\documentstyle [prl,aps]{revtex}
\input epsf
\begin{document}

\draft
\title{First Characterization of the Ultra-Shielded Chamber
in the Low-noise Underground Laboratory (LSBB)
of Rustrel Pays d'Apt}

\author{
G. Waysand$^{1,4,*}$, D. Bloyet$^{2}$, J.P. Bongiraud$^{3}$, 
J.I. Collar$^{1,4}$, C. Dolabdjian$^{2}$,
Ph. Le Thiec$^{3}$
}
\address{ 
$^{1}$Groupe de Physique des Solides (UMR CNRS 75-88), Universit\'es Paris 
6 \& 7, 
2 Pl. Jussieu, Paris 75251, France\\
$^{2}$Groupe de Recherche en Informatique, Image et Instrumentation 
de Caen ISMRA, 
Universit\'e de Caen, 14050 Caen, France\\
$^{3}$Laboratoire de Magn\'etisme du Navire, ENSEIG/INPG, 
Universit\'e Joseph Fourier, 38 Grenoble, France\\
$^{4}$Laboratoire Souterrain \`a Bas Bruit (LSBB) 
de Rustrel-Pays d'Apt, 84400 Rustrel, France\\
}
\wideabs{
\maketitle
\begin{abstract}
\widetext
In compliance with international agreements on nuclear weapons limitation, 
the French ground-based nuclear arsenal has been decommissioned in its 
totality. One of its former underground missile control centers, located 
in Rustrel, 60 km east of Avignon (Provence) has been converted into 
the ``Laboratoire Souterrain \`a Bas Bruit de Rustrel-Pays d'Apt'' (LSBB). 
The deepest experimental hall (500 m of calcite rock overburden) 
includes a 100 m$^{2}$  area of sturdy flooring suspended by and resting on 
shock absorbers, entirely enclosed in a 28 m-long, 8 m-diameter, 1 cm-thick 
steel Faraday cage. This results in an unparalleled combination of 
shielding against cosmic rays, acoustic, seismic and electromagnetic 
noise, which can be exploited for rare event searches using ultra 
low-temperature and superconducting detectors. The first 
characterization measurements in this unique civilian site 
are reported. {\tt http://home.cern.ch/collar/RUSTREL/rustrel.html}
\end{abstract}

\pacs{
$^{*}$E-mail: waysand@gps.jussieu.fr}
}

\narrowtext
{\bf 1. Description of the infrastructure}\\

Rustrel, a small village in the Pays d'Apt, is one hour by 
car east of Avignon, in the heart of Provence (southeast France). 
The high-speed train connection from Paris to Avignon takes 
three hours and twenty minutes. Marseille-Marignagne is the 
nearest international airport, one hour and a half from Rustrel. 
The village is just south of the Plateau d'Albion, the former 
location for the ground-based component of the French air 
force's ``force de frappe'' (strike force). The military selection 
of this location was presumably due to the proximity of the 
eastern French border, low density of population, good rock 
quality, and the existence of mountains offering natural 
protection. Missile silos were spread over a large area, 
with two underground launching centers on both sides 
of the Mount Ventoux.

One of these centers has been spared from destruction 
and is now converted into a laboratory (LSBB); the usable 
spare parts and technical subsystems from the second 
center have been preserved for future maintenance of the LSBB. 
The LSBB consists of 3.2 km of reinforced concrete galleries 
below the "Grande Montagne" (1,010 m at the summit), 
joining various halls.  An heliport is available in front 
of the entrance area, which houses office space and living 
quarters. A telecommunications area in the summit is directly 
connected via optical fiber to the deepest part of the galleries, 
the launching control room. 

Two major experimental halls are available; the shallowest 
(350 m$^{2}$, 7 m ceiling height, 50 m rock overburden) is located 400 
meters from the entrance and is shielded in the same manner as 
the launching control room (one kilometer further down the corridors, 
protected 
by 500 m of calcite rock). This control room is the second hall 
of interest and includes 100 m$^{2}$ of sturdy flooring suspended 
by shock absorbers. The room is entirely surrounded by a horizontal 
steel capsule 28 m-long, 8 m in diameter and 1 cm-thick; entrance 
doors are clamped by electrical contacts to ensure the sealing of 
this peculiar Faraday cage. Several auxiliary galleries at the same depth can 
be used for experiments not needing exceptional EM shielding.  

Due to its previous purpose, the whole setup was designed to 
offer maximum safety against intrusion and nuclear attack: External 
and internal steel doors can bear 20 bars of overpressure and 
emergency generators respond in a tenth of a second to energy 
supply perturbations. Ten optical fibers are available for 
telecommunications with the exterior. The whole area is fully 
air-conditioned, with ventilation and running water reaching the 
deepest hall. Road traffic is very scarce 
within two kilometers from the laboratory.\\

{\bf 2. Radiation shielding \& natural radioactivity}\\

{\bf 2.1 Neutrons}\\
The capsule and surrounding rooms are located at a depth of $\sim$1,500 
meters of water equivalent (m.w.e.). In this sense, LSBB 
ranks about average when compared to other underground laboratories. 
Nevertheless, this depth is more than enough to ensure screening 
of secondary cosmic neutrons (fig. 1). Indeed, neutrons produced 
by natural radioactivity in the surrounding rocks are dominant 
below few hundreds of m.w.e., independent of the nature of the 
shielding materials used in the experiments 
(deeply-reaching muons can produce neutrons in this shielding). 
An increased depth brings no further reduction in neutron flux 
\cite{sheff}.
\begin{figure}[tbp]
\epsfxsize = \hsize \epsfbox{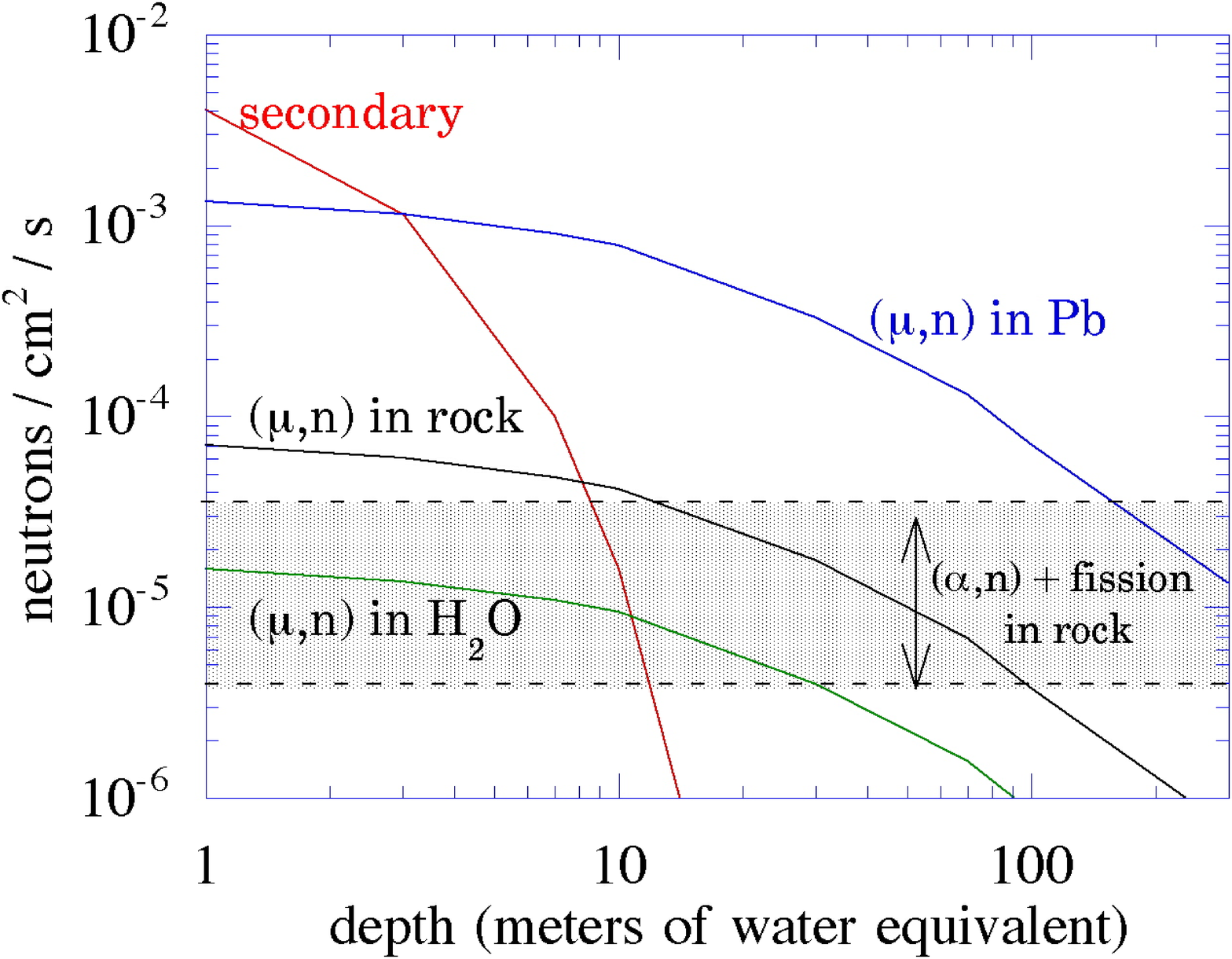}
\caption{Components of neutron flux as a function of 
underground depth. The greyed band covers the typical contribution 
from natural radioactivity in rocks, based on large-depth measurements 
and the average U and Th concentration in the Earth's crust. 
The contribution to the flux from muon interactions is derived 
from the systematics and measurements of [2] (similar results 
are obtained in [3]). } 
\end{figure}

{\bf 2.2 Muons}\\
At 1,500 m.w.e. the muon flux becomes a second order concern for most 
of the activities envisioned at LSBB, and can be further suppressed 
by anticoincidence with an active veto (plastic scintillator) without 
creating a substantial dead time. In the case of ultra-low temperature 
experiments, long-lived heating by passing muons 
(a problem at ground level) is a rare occurrence at a flux of 
$\sim 5\cdot 10^{-3} \mu/m^{2}/s$ (fig. 2).
\begin{figure}[tbp]
\epsfxsize = \hsize \epsfbox{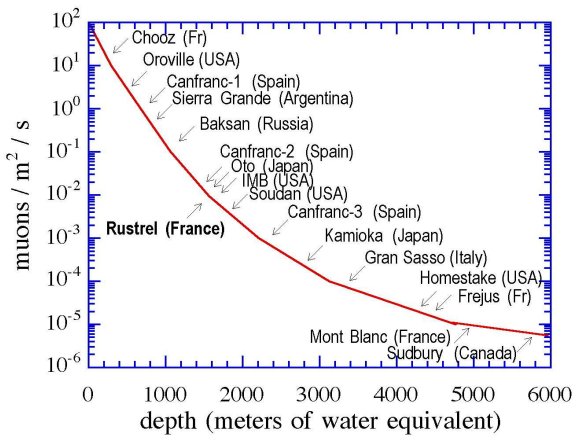}
\caption{Muon flux as a function of depth for several underground 
laboratories [4].} 
\end{figure}

{\bf 2.3 Rock radioactivity}\\
 Rock samples were extracted from exposed walls in secondary galleries, 
measured by the CERN radioprotection service and compared with a 
reference calcite rock from a low-depth gallery nearby Paris. 
Only two isotopes were detected above background 
(table 1).
\begin{center}
\begin{tabular}[t]{p{2.0 truecm}p{3.5 truecm}p{2.0 truecm}}
\hline
Isotope & Boissise la Bertrand  &  LSBB \\
 & (Seine et Marne)	 &  (Rustrel) \\
  & & \\
$^{137}$Cs & 0.204 Bq & 0.437 Bq \\
$^{226}$Ra & 2.030 Bq & 0.645 Bq \\
\hline
\end{tabular}
\end{center}
{\small Table 1: Identifiable rock radioactivity in Boissise 
(near Paris), and in the Laboratoire Souterrain \`a Bas Bruit 
(LSBB).}\\

Comparable levels were found, within the factor 2 uncertainty 
in the activities. While this is nowhere close to an exhaustive 
measurement, it allows to discard the possibility that Rustrel 
rock might be unusually ``hot''.\\

{\bf 2.4 Airborne radon}\\
The radon concentration in the atmosphere of the capsule 
was measured during three weeks in January 1998 using a RAD7 NITON 
radonmeter (a self-contained continuous-monitoring solid state 
alpha detector \cite{niton}). An average value of 28 Bq/m$^{3}$ 
(0.77 pCi/l) was obtained \cite{rn}. As a reference, this can be compared 
with the maximum acceptable 45 Bq/m$^{3}$ in US households, or with the 
lowest values achieved in the Gran Sasso underground 
laboratory, 20-50 Bq/m $^{3}$ \cite{gsasso}. Similar results have 
been obtained in April 1999. This low rate, comparable to outdoor 
measurements in the area, is a factor 15 lower than during military 
operation. The improvement was obtained by opening the vertical 
escape chimney of the site, allowing for natural ventilation, and 
by turning off a cooling unit within the capsule. The escape 
chimney was normally obstructed for security reasons; as a result, 
in spite of the strong ventilation in the gallery, the deepest hall 
was almost a dead end for air circulation.\\

{\bf 3. Seismicity and acoustic noise}\\

The area of Rustrel is well-studied from a geological point of view.  
It is at the center of a 30 km circle free of active faults in spite 
of the relative proximity to the Alps. This structural configuration 
is the reason for the absence of local seismicity during the last 
eleven hundred years \cite{seism}.
The acoustic environmental noise has been roughly characterized in 
a 20 m$^{2}$ shielded room adjacent to the capsule during the 
installation of the SIMPLE experiment \cite{simple}, which relies 
precisely on acoustic detection of the weak signal arising from 
the vaporization of superheated freon droplets suspended 
in a gel matrix. At the level of sensitivity of our room monitoring 
microphones, no significant activity was detected after the 
ventilation ducts were muffled.\\

{\bf 4. Electromagnetic shielding}\\

{\bf 4.1 A unique underground shielding}\\
The peculiar shielding of the main experimental halls was designed 
with the intention of protecting electronic equipment from the huge 
electromagnetic pulse created by a nearby nuclear explosion. 
This is the reason why, instead of a conventional Faraday cage made 
of thin copper, the choice was made for a thick (1 cm) steel shielding. 
As a result, these large cages attenuate not only high-frequency 
electromagnetic waves but also low-frequency and even DC 
(e.g., the magnetic field of the Earth).  Due to their large 
dimensions it is possible to locally create large magnetic fields 
as long as the walls have not reached their magnetic saturation. 
In other words, within these cages one can have very low magnetic 
fluctuations even at non-zero magnetic field values.\\ 

{\bf 4.2  DC domain}\\
The chosen steel was not optimized for magnetic shielding; nevertheless, 
the residual magnetic field inside the capsule is lower than 6 $\mu$T 
(compare with 46 $\mu$T for the Earth's magnetic field at the LSBB 
latitude). The measurements were done with triaxial fluxgate 
magnetometers with a bandwidth from 0 to 5 Hertz, a noise level of 0,5 nT 
(peak to peak) and an absolute precision of 200 nT (0.2 $\mu$T) . Over 
a period longer than 12 hours a remarkable long-term stability and 
low noise level (less than 20 nT) were observed.
This performance is not extreme, yet very impressive when observed 
over such a large experimental area (one would have to wait more than 
12 hours to observe, in a square loop of $0.3\times0.3 ~mm^{2}$, 
a magnetic flux variation larger than one quantum flux!). 
Such a long-term magnetic stability allows for the utilization 
of SQUID detectors with large pick-up coils.
A Hall-effect gaussmeter applied directly on the steel walls 
just above the welding lines revealed only a weak local magnetization 
(local magnetic field smaller than 100 $\mu$T).
At the expense of a few precautions (displacement of the ventilating 
units, compensated AC wiring) the EM quality of the site can be 
improved even further.\\

{\bf 4.3 Dynamic fluctuations}\\
Using a triaxial fluxgate connected to a spectrum analyzer, 
the performance of the shielding was measured from 1 to 1000 Hz. 
No detectable signal above the noise level of the measuring chain 
was obtained, indicating that the magnetic fluctuations are 
lower than 2.5 pT/Hz$^{1/2}$. Finally, in early August 1999 
a high-Tc SQUID was operated for a short period inside the capsule; 
from this measurement it was concluded that in the same frequency 
range the noise level is below 600 fT/Hz$^{1/2}$.\\

{\bf 5. Conclusion}\\

This first set of characterization measurements indicates that 
a singular combination of shielding features makes of LSBB 
a site of choice for low-noise experiments in the fields of 
ultra-low temperature physics, superconductivity, biology, 
metrology and astroparticle physics.\\

{\it Acknowledgements:} 
We are indebted to M. Auguste, G. Boyer, A. Cavaillou and 
L. Ibtiouene for their help in performing these measurements.


\begin{thebibliography}{999}
\bibitem{sheff} J.I. Collar, T.A. Girard, D. Limagne and G. Waysand, 
Procs. of the 1st International Workshop on the Identification of 
Dark Matter (IDM96), Sheffield, U.K., 1996. N.Spooner ed., 
World Scientific (Singapore) {\tt astro-ph/9610266}.
\bibitem{gors} G.V. Gorshov \emph{et al.}, 
Sov. J. Nucl. Phys. {\bf 13} (1971) 450. 
\bibitem{heu} G. Heusser, Nucl. Instr. Meth {\bf A369} (1996) 539. 
\bibitem{angel} A.Morales, Inaugural Lecture of the 95-96 academic year, 
Universidad de Zaragoza.
\bibitem{niton}Niton electronics, P.O. Box 368, Bedford MA 01730, USA.
\bibitem{rn}{\tt http://home.cern.ch/collar/RUSTREL/radon.JPG}
\bibitem{gsasso}C. Arpesella \emph{et al.}, Health Phys. {\bf 72} 
(1997) 629.
\bibitem{seism}B. Grellet, Ph. Combes and D. Carbon, 
Geo-Ter Laboratory, Universit\'e de Montpellier: 
Procs. of the 1997 meeting of the ANDRA 
(French national agency for nuclear waste), 
Bagnols-sur-C\`eze, 20-21 October 1997. Published by ANDRA.
\bibitem{simple}J.I. Collar, T.A. Girard, D. Limagne, H.S. Miley, 
J. Puibasset and G. Waysand,
Procs. of the 2nd Intl. Workshop on the Identification of Dark Matter (IDM98), 
Buxton, U.K., 1998. N.Spooner ed., World Scientific (Singapore); 
{\tt http://taup99.in2p3.fr/cgi-bin/taup/ 
get\_scan.pl?Monday+workshop+J\_Collar}.
\end{thebibliography}
\end{document}